\documentclass[preprint,12pt]{elsarticle}

\usepackage{graphicx}
\usepackage{amsmath}
\usepackage{amssymb}
\usepackage{afterpage}
\usepackage{graphics}
\usepackage{float}
\usepackage{rotating}
\usepackage{xspace}
\usepackage{dcolumn}
\usepackage{bm} 

\journal{Astroparticle Physics}

\begin{document}

\begin{frontmatter}



\title{The Search for Neutrino-Antineutrino Mixing Resulting from Lorentz Invariance Violation using neutrino interactions in MINOS\tnoteref{t1}}
\tnotetext[t1]{FERMILAB-PUB-13-025-E}

\author[label1]{B.~Rebel }
\ead{brebel@fnal.gov}
\address[label1]{Fermi National Accelerator Laboratory, Batavia, Illinois 60510, USA}
\author[label2]{S.~Mufson}
\ead{mufson@astro.indiana.edu}
\address[label2]{Indiana University, Bloomington, Indiana 47405, USA}

\date{\today}          

\begin{abstract}
We searched for a sidereal modulation in the rate of neutrinos produced by the NuMI beam and observed by the MINOS far detector.  The detection of such harmonic signals could be a signature of neutrino-antineutrino mixing due to Lorentz and CPT violation as described by the Standard-Model Extension framework.  We found no evidence for these sidereal signals and we placed limits on the coefficients in this theory describing the effect.  This is the first report of limits on these neutrino-antineutrino mixing coefficients.  
\end{abstract}


\end{frontmatter}

\newpage


\section{Introduction}
\label{sect:intro}

The sensitive interferometric nature of neutrino oscillations offer interesting and novel ways to search for Lorentz invariance violation (LV) and CPT violation (CPTV) that may arise from physics at the Planck scale, $10^{19}$~GeV.  The SME is effective field theory that contains the Standard Model, General Relativity, and all possible effective operators for Lorentz violation~\cite{CKcomb}.  In the SME, LV and CPTV could manifest themselves at observable energies through a dependence of the neutrino oscillation probability on the direction of neutrino propagation with respect to the Sun-centered inertial frame.   An experiment that has both its neutrino beam and detector fixed on the Earth's surface could then observe a sidereal modulation in the number of neutrinos detected from the beam. 

MINOS is such an experiment~\cite{Michael:2008bc}.  It uses Fermilab's NuMI neutrino beam~\cite{numi} and two neutrino detectors -- a near detector (ND) that is 1.04~km from the beam target and a far detector (FD) that is 735~km from the beam target.  Both detectors are magnetized to approximately 1.4~T, allowing for the identification of $\mu^{-}(\mu^{+})$ produced in charged-current (CC) neutrino (antineutrino) interactions.  In a series of recent papers \cite{ndlv, paper2, paper3}, MINOS searched for LV and CPT violation with neutrino and antineutrino data recorded in its detectors~\cite{Adamson:2011ig1,Adamson:2011ig2}.  These analyses focused on detecting oscillation signals in neutrino and antineutrino flavor-changing reactions that vary with a sidereal period, as predicted by the Standard Model Extension (SME).  MINOS found no evidence for LV and CPTV in these analyses.  In the present paper, we extend the search for LV and CPTV using the timestamps of the neutrino data made available by the MINOS collaboration to look for sidereal signals that would indicate neutrinos changing to antineutrinos.  Although there have been many searches for LV and CPTV with the SME~\cite{tables}, this is the first to probe the possibility of neutrino-antineutrino mixing.   

Denote the usual neutrino survival probability in the two-flavor approximation as $P^{(0)}_{\nu_\mu \rightarrow \nu_\mu} \approx 1-\sin^2{(2 \theta_{23})} \sin^2{(1.27 \Delta m^2_{32} L /E)}$, where $\theta_{23}$ is the angle describing mixing between the second and third mass states and $\Delta m^{2}_{32}$ is the difference in the squares of those mass states.  The energy of the neutrino is $E$ and the distance it travels is $L$.  Then LV and CPTV  that would cause mixing between neutrinos and antineutrinos introduces an additional perturbation term to the survival probability~\cite{dkm},
\begin{equation}
\label{eq:totalprob}
P_{\nu_\mu \rightarrow \nu_\mu} = P^{(0)}_{\nu_\mu \rightarrow \nu_\mu} + P^{(2)}_{\nu_\mu \rightarrow \nu_\mu},
\end{equation}
where the perturbation term $P^{(2)}_{\nu_\mu \rightarrow \nu_\mu}$ can be written
\begin{eqnarray}
\label{eq:osc}
P^{(2)}_{\nu_\mu \rightarrow \nu_\mu} &=& L^2 \biggl\{P_{\mathcal{C}} 
 +P_{\mathcal{B}_{s}}\sin2\omega_{\oplus}T_{\oplus} + P_{\mathcal{B}_{c}}\cos2\omega_{\oplus}T_{\oplus} \\
&+& P_{\mathcal{F}_{s}}\sin4\omega_{\oplus}T_{\oplus} + P_{\mathcal{F}_{c}}\cos4\omega_{\oplus}T_{\oplus} 
\biggr\}.\nonumber 
\end{eqnarray}
Here $L = 735$~km is the distance from neutrino production in the NuMI beam to the MINOS FD~\cite{minoscc}, $\omega_\oplus= 2 \pi/(23^h 56^m 04.0982^s)$ is the Earth's sidereal frequency, and $T_\oplus$ is the local sidereal arrival time of the neutrino event.  The parameters $P_{\mathcal{C}}$, $P_{\mathcal{B}_{s}}$, $P_{\mathcal{B}_{c}}$, $P_{\mathcal{F}_{s}}$, and $P_{\mathcal{F}_{c}}$ contain the LV and CPTV information on neutrino-antineutrino mixing.  They depend on the SME coefficients $\tilde H^{\,\alpha}_{a \bar b}$ and $\tilde g^{\,\alpha\beta}_{a \bar b}$, the neutrino energy, and the direction of the neutrino propagation in a coordinate system fixed on the rotating Earth~\cite{dkm}.   As is clear from Eq.(\ref{eq:osc}), we need only the even harmonics $2\omega_{\oplus}T_{\oplus}$ and $4\omega_{\oplus}T_{\oplus}$ for this analysis. Further, since we are testing the theory with a harmonic analysis, there is no sensitivity to $P_{\mathcal{C}}$.

In the minimal SME, neutrino-antineutrino mixing $\nu_a \leftrightarrow \bar \nu_b$ is controlled by the complex coefficients  $\tilde H^{\,\alpha}_{a \bar b}$ and $\tilde g^{\,\alpha\beta}_{a \bar b}$ ($a = e, \mu, \tau$ and $\bar b = \bar e, \bar \mu, \bar \tau$).  Sidereal modulations detectable by MINOS are produced by 6 of the $\tilde H^{\,\alpha}_{a \bar b}$ coefficients and 60 of the  $\tilde g^{\,\alpha\beta}_{a \bar b}$ coefficients~\cite{dkm}.  For the $\tilde H^{\,\alpha}_{a \bar b}$ coefficients, only three independent pairs of  flavor subscripts ($a \bar b = e \bar\mu, e \bar\tau, \mu \bar \tau$) and two spatial components ($\alpha =$ {\it X, Y}) produce sidereal variations.  For the  $\tilde g^{\,\alpha\beta}_{a \bar b}$ coefficients, MINOS can access six pairs of flavor indices ($a \bar b = e \bar e, e \bar\mu, e \bar\tau, \mu \bar \mu, \mu \bar \tau, \tau \bar \tau$) and 10 different spacetime components ($\alpha\beta =$ {\it XT, YT, XZ, YZ, ZX, ZY, XX, YY, XY, YX}).  

For the neutrino propagation direction, we specify the direction vectors in a right-handed coordinate system defined by the colatitude of the NuMI beam line $\chi = (90^\circ -$ latitude) = 42.17973347$^{\circ}$, the beam zenith angle $\theta=86.7255^\circ$, measured from the $z$-axis and which points toward the local zenith, and the beam azimuthal angle, $\phi=203.909^\circ$, measured counterclockwise from the $x$-axis, which points south.  

\section{Data Analysis}
\label{sect:dataAnalysis}

This analysis uses a data set of neutrino interactions acquired from May, 2005 through April, 2012~\cite{Adamson:2011ig1,Adamson:2011ig2} when the beam was running in a mode that produces 91.7\% $\nu_{\mu}$, 7\% $\overline{\nu}_{\mu}$, and 1.3\% $\nu_e+ \overline{\nu}_{e}$~\cite{Adamson:2011qu}.  
The interactions were selected using standard MINOS criteria for beam and data quality.  In addition, the events were required to interact within the 4.2~kiloton FD fiducial volume.  This selection enables MINOS to establish each event as a CC $\nu_\mu$ interaction by identifying the outgoing $\mu^-$.  The method is described more fully in~\cite{Adamson:2011ig1,Adamson:2011ig2}. As in~\cite{ndlv, paper2, paper3}, we focused on CC events to maximize the $\nu_\mu$ disappearance signal.  There are a total of 2,463 CC events in this analysis.  

The analysis proceeded by first tagging each neutrino interaction with the local sidereal time (LST) of its spill -- the GPS time of the spill, accurate to 200~ns~\cite{minosToF}, converted to sidereal time.  The event's LST was then converted to local sidereal phase, LSP = LST$\times (\omega_\oplus/2\pi)$, a parameter in the range 0-1.  Event times were not corrected for their time within a 10~$\mu$s spill, an approximation that introduces no significant systematic error into the analysis.  The data for each spill were binned into two histograms.  In one histogram, we added any interaction from each spill; in the second histrogram we added the number of protons impinging on the target (POT) for each spill.  These two histograms have 32 bins in LSP because the Fast Fourier Transform (FFT) algorithm we use to search for sidereal variations~\cite{numrec} works most efficiently for $2^{\mathcal N}$ bins, where ${\mathcal N}-1$ is the number of harmonics  to be computed in addition to a constant term.   Since the highest harmonic in Eq.~(\ref{eq:osc}) is $4\omega_{\oplus}T_{\oplus}$, we adopted ${\mathcal N} = 5$ as the binning.   Each phase bin spans 0.031 in LSP or 45 minutes in sidereal time.

\begin{figure}[h]
\centerline{\includegraphics[width=3.25in]{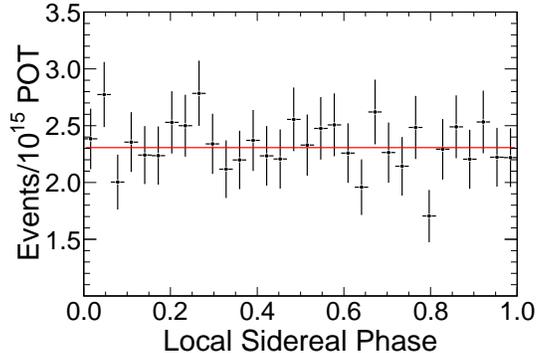}}
\caption{\label{fig:data_rate} The phase diagram of the CC neutrino event rate for the FD data.  The mean rate of $2.31 \pm 0.05$ events per $10^{15}$~POT is superposed.  For the fit, $\chi^{2}/ndf = 23.7/31$.}
\end{figure}

After adding all the events and POT in the data set into the two histograms, we divide them to get the rate of interactions per POT,  ($\nu_\mu$~CC-events)/(POT) as a function of LSP.  This histogram is the one used to search for sidereal modulations in the neutrino rate.  Figure~\ref{fig:data_rate} shows this histogram for our analysis.
The statistically significant fit to a constant rate implies there are no sidereal modulations in the data sample.  

We next performed an FFT on the rate histogram in Fig.~\ref{fig:data_rate} and computed the power in the harmonic terms $2\omega_{\oplus}T_{\oplus}$ and $4\omega_{\oplus}T_{\oplus}$ appearing in the oscillation probability, Eq.(\ref{eq:osc}).   Let $S_2$ be the power returned by the FFT for the second harmonic term  $\sin{(2\omega_{\oplus}T_{\oplus}})$ and  $C_2$ be the power returned for the second harmonic term  $\cos{(2\omega_{\oplus}T_{\oplus}})$; similarly define $S_4$ and $C_4$. Then the statistics we used in our search are
\begin{equation}
p_2 = \sqrt{S_2^2 + C_2^2}, \text{ and } p_4 = \sqrt{S_4^2 + C_4^2}.
\end{equation}
We added the powers in quadrature to eliminate the effect of the arbitrary choice of a zero point in phase at $0^h$ LST.  The results of the FFT analysis are given in Table~\ref{table:dataPower}.  Although only the even harmonics $p_2$ and $p_4$ are needed for this analysis, we include the power in the odd harmonics $p_1$ and $p_3$ as a check on whether alternative theories on LV that put power into these modes are realizable. 
\begin{table}[h]
\caption{\label{table:dataPower} Results for the $p_1, \dots, p_4$ statistics for the histogram shown in Fig.~\ref{fig:data_rate}.  The third column gives the probability, $\cal{P}_F$, that the measured power is due to a noise fluctuation. }
\begin{center}
\begin{tabular}{|c|c|c|}
\hline\hline
~~Statistic~~ & ~~$p$(FFT)~~& ~~~~~$\cal{P}_F$~~~~~ \\ 
\hline
$p_1$ &  0.928 & 0.65\\ 
$p_2$  & 0.574 & 0.89 \\
$p_3$  & 1.388 & 0.48 \\
$p_4$  & 1.223 & 0.53 \\
\hline \hline
\end{tabular}
\end{center}
\end{table}

As described in~\cite{paper2}, we determined the statistical significance of the harmonic powers $p$(FFT) in Table~\ref{table:dataPower} by using the data themselves to construct $10^{4}$ simulated experiments without a sidereal signal.  To construct a simulated experiment we used the sidereal time distribution of the beam spills to determine a new, randomly selected LSP for each interaction in the data.  Using this randomly chosen phase, we added each interaction to one histogram.  Similarly, for each spill we selected a random LSP and added the number of POT in the spill to a second histogram.  The FD records fewer than one interaction per spill, so the spill and interaction LSP for these simulated experiments can be selected independently without introducing any biases in the procedure.  As for the real neutrino data set, the histograms had 32 bins in LSP.  We then divided the events histogram by the POT histogram to obtain the rate histogram for the simulated experiment.  We repeated this process $10^4$ times.  This procedure ensures that the spill times are distributed properly in LSP.  Most importantly, it scrambles any coherent signal, if present, in the data and eliminates it in the simulated experiments.  

\begin{figure}[h]
\centerline{\includegraphics[width=3.25in]{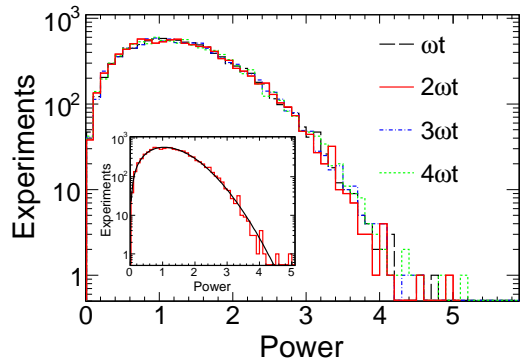}}
\caption{\label{fig:mc_power} The distributions for the quadratic sum of powers  $p_1, \ldots, p_4$ from the FFT analysis of $10^{4}$ simulated experiments without a sidereal signal.  The inset shows the distribution for $p_{2}$ with a fit to a Rayleigh distribution with $\sigma = 1.49$ superposed.
}
\end{figure}

We next performed an FFT on each simulated histogram of (events/POT) and computed the power in the four harmonic terms ($\omega_{\oplus}T_{\oplus}, \ldots, 4\omega_{\oplus}T_{\oplus}$).    Fig.~\ref{fig:mc_power} shows the distribution of $p_1, \ldots, p_4$ for the $10^{4}$ simulated experiments.  
The distributions for  $p_1, \ldots, p_4$ are quite similar over the full range of observed powers.  As seen in the inset to Fig.~\ref{fig:mc_power},
the $p_{2}$ distribution is well described by a Rayleigh distribution, 
as expected in these simulated experiments since they were constructed so that the power in the sine and cosine terms are uncorrelated and normally distributed.  

We assume that the threshold for signal detection in any harmonic is the quadratic power $p(\text{FFT})$ that is greater than 99.7\% of the entries in its respective $p_1, \ldots, p_4$ histogram.  The values of $p_{99.7\%}(\text{FFT})$ for all four harmonics lie between 3.40 -- 3.50, with $p_2 = p_4 = 3.45$.  We adopt $p_{99.7\%}(\text{FFT}) = 3.45$  as the 99.7\% confidence level (C.L.) for the probability that a measured quadratic sum of powers for any harmonic was {\it not} drawn from a distribution having a sidereal signal.    As neither $p_2$ nor  $p_4$ exceed our detection threshold, we conclude that there is no evidence for a sidereal modulation resulting from neutrino-antineutrino mixing as predicted by Eq.(\ref{eq:osc}) in the neutrino data set.  This result is consistent with the conclusion that the constant rate fit to the data in Fig.~\ref{fig:data_rate} implies there is no sidereal modulation in the data.

We determined the minimum detectable sidereal modulation for this analysis by injecting a sidereal signal of the form $A\sin(\omega_{\oplus} T)$, where $A$ is a fraction of the mean event rate, into a new set of $10^{4}$ simulated experiments and repeating the FFT analysis.  We found that every experiment gave $p \geq 3.45$, or exceeded the detection threshold, when $A = 13\%$ of the mean rate.

In Table~\ref{table:dataPower} we also give $\cal{P}_{F}$, the probability that the power measured in a particular harmonic is due to a noise fluctuation.  In other words, $\cal{P}_F$ is the probability of drawing a value at least as large as found for $p_1, \dots,  p_4$ from the parent distribution in Fig.~\ref{fig:mc_power} of the powers for numerical experiments without a sidereal signal.  The large values for $\cal{P}_F$ also make the case that the powers we measure are the result of statistical fluctuations.

We investigated the sensitivity of our results to several sources of systematic uncertainties.  In the previous MINOS analyses~\cite{ndlv, paper2}, the degradation of the NuMI target was observed to cause a drop in the event rate.   Further, the NuMI target was replaced multiple times during the data taking period of this analysis which could also have affected the event rate.  We investigated how these changes in the event rate would affect the detection thresholds and found this source of systematic uncertainty to be negligible.  
 
Day-night effects are another potential source of systematic uncertainty for a data set like ours that does not have uniform coverage throughout a solar year.  We find the mean rate during the day to be $2.32 \pm 0.06$ events/$10^{18}$~POT and the mean rate at night to be $ 2.21 \pm 0.07$ events/$10^{18}$~POT.  Since there is no statistically significant difference in these rates, we conclude that diurnal effects cannot be masking a true sidereal signal in the data.  

There is a known uncertainty of $\pm 1$\% in the recorded number of POT per spill~\cite{Adamson:2007gu}.  We verified this uncertainty could not introduce a modulation that would mask a sidereal signal as done in~\cite{paper3}.  We further determined that long term drifts in the calibration of the POT recording toroids of the size $\pm 5$\% over six months would not alter the detection threshold.  We thus conclude that POT counting uncertainties are not masking a sidereal signal.  

Compton and Getting~\cite{CG} first pointed out that atmospheric effects can mimic a sidereal modulation in the event rate if there were a solar diurnal modulation  that beats with a yearly modulation.  As in~\cite{paper3}, we found the amplitude of the potential faux sidereal modulation would be only $1.6\%$ of our minimum detectable modulation and therefore would not mask a sidereal signal in these data.

\section{Limits}
\label{sect:limits}

We determined the confidence limit for a particular SME coefficient by simulating a set of experiments in which we inject an LV signal into a Monte Carlo simulation that does not include a sidereal modulation.  In this simulation, neutrinos are generated by modeling the NuMI beam line, including hadron production by the 120 GeV$/c$ protons striking the target and the propagation of the hadrons through the focusing elements and decay pipe to the beam absorber.  It then calculates the probability that any neutrinos generated traverse the FD.  The FD neutrino event simulation takes the neutrinos from the NuMI simulation, along with weights determined by decay kinematics, and uses this information as input into the simulation of the interactions in the FD.  We inject a sidereal signal in the simulation by calculating the survival probability for each simulated neutrino based on Eq.~(\ref{eq:totalprob}) using a chosen value for the magnitude of the SME coefficient, the energy of the simulated neutrino, and the distance the neutrino travels to the FD.  We start by setting the SME coefficient to zero and then continually increasing its magnitude until either $p_2$ or $p_4$ crosses the detection threshold of 3.45.  We took this value of the SME coefficient to be its 99.7\% C.L. upper limit.  The SME coefficients can be either positive or negative, so this limit is the lower limit on the SME coefficient in the case it is negative.  We repeated the simulation for this coefficient 250 times and averaged  the resulting value of the coefficient.  Each of the simulated experiments contained the same number of interactions as the data set.  

The 99.7\% C.L. limits on the SME coefficients are given in Table~\ref{table:limits1} and Table~\ref{table:limits2}. 
\begin{table}[h]
\caption{\label{table:limits1} Upper limits to the modulus of the ${\mathcal Re}$ parts of the $\tilde H^{\,\alpha}_{a \bar b}$ coefficients in  [GeV] describing neutrino-antineutrino mixing in the SME theory.  The limits were determined from the 99.7\% threshold for a positive sidereal signal in the neutrino rates in the MINOS FD.  The ${\mathcal Re}$ and ${\mathcal Im}$ components of these SME coefficients are computed identically and are equal.} 

\begin{center}
\begin{tabular}{|c|c|c|c| } 
\hline \hline 

$\alpha$ & ${|\mathcal Re}(\tilde H^{\alpha}_{e \bar \mu})|$ & 
$|{\mathcal Re}(\tilde H^{\alpha}_{e \bar \tau} )|$  & 
$|{\mathcal Re}( \tilde H^{\alpha}_{\mu \bar \tau} )|$ \\

\hline \hline

{\it X} &$3.3\times10^{-21}$&$3.4\times10^{-21}$&$1.1\times10^{-22}$\\
{\it Y} &$3.3\times10^{-21}$&$3.6\times10^{-21}$&$1.0\times10^{-22}$\\

\hline \hline

\end{tabular}\\
\end{center}
\end{table}
\begin{table}[h]
\caption{\label{table:limits2} Upper limits to the modulus of the ${\mathcal Re}$ parts of the $\tilde g^{\,\alpha\beta}_{a \bar b}$ coefficients describing neutrino-antineutrino mixing in the SME theory.  The limits were determined from the 99.7\% threshold for a positive sidereal signal in the neutrino rates in the MINOS FD.  The ${\mathcal Re}$ and ${\mathcal Im}$ components of these SME coefficients are computed identically and are equal.}

\begin{center}
\begin{tabular}{|c|c|c|c|c|c|c| } 
\hline \hline

$\alpha\beta$ & $|{\mathcal Re}(\tilde g^{\,\alpha\beta}_{e \bar e} )|$ & 
$|{\mathcal Re}(\tilde g^{\,\alpha\beta}_{e \bar \mu})|$  & 
$|{\mathcal Re}(\tilde g^{\,\alpha\beta}_{e \bar \tau})|$  & 
$|{\mathcal Re}(\tilde g^{ \,\alpha\beta}_{\mu \bar \mu} )|$  & 
$|{\mathcal Re}(\tilde g^{\,\alpha\beta}_{\mu \bar \tau} )|$  & 
$|{\mathcal Re}(\tilde g^{\,\alpha\beta}_{\tau \bar \tau} )|$ \\

\hline \hline

{\it XT} &$7.6\times10^{-22}$&$7.6\times10^{-22}$&$8.2\times10^{-22}$&$8.9\times10^{-24}$&$8.6\times10^{-24}$&$1.8\times10^{-22}$\\
{\it YT} &$7.6\times10^{-22}$&$7.6\times10^{-22}$&$7.6\times10^{-22}$&$8.6\times10^{-24}$&$8.4\times10^{-24}$&$1.8\times10^{-22}$\\
{\it XZ} &$1.2\times10^{-21}$&$1.2\times10^{-21}$&$1.2\times10^{-21}$&$1.3\times10^{-23}$&$1.3\times10^{-23}$&$2.9\times10^{-22}$\\
{\it YZ} &$1.2\times10^{-21}$&$1.2\times10^{-21}$&$1.2\times10^{-21}$&$1.3\times10^{-23}$&$1.4\times10^{-23}$&$2.9\times10^{-22}$\\
{\it ZX} &$1.0\times10^{-21}$&$1.0\times10^{-21}$&$1.0\times10^{-21}$&$1.2\times10^{-23}$&$1.1\times10^{-23}$&$2.4\times10^{-22}$\\
{\it ZY} &$1.0\times10^{-21}$&$1.0\times10^{-21}$&$1.0\times10^{-21}$&$1.1\times10^{-23}$&$1.2\times10^{-23}$&$2.4\times10^{-22}$\\
{\it XX} &$2.0\times10^{-21}$&$2.0\times10^{-21}$&$2.0\times10^{-21}$&$2.3\times10^{-23}$&$2.3\times10^{-23}$&$4.8\times10^{-22}$\\
{\it YY} &$2.0\times10^{-21}$&$2.1\times10^{-21}$&$2.1\times10^{-21}$&$2.2\times10^{-23}$&$2.2\times10^{-23}$&$4.8\times10^{-22}$\\
{\it XY} &$2.0\times10^{-21}$&$2.0\times10^{-21}$&$2.1\times10^{-21}$&$2.3\times10^{-23}$&$2.3\times10^{-23}$&$4.8\times10^{-22}$\\
{\it YX} &$2.0\times10^{-21}$&$2.0\times10^{-21}$&$2.0\times10^{-21}$&$2.2\times10^{-23}$&$2.2\times10^{-23}$&$4.8\times10^{-22}$\\

\hline \hline

\end{tabular}
\end{center}
\end{table}
By setting all but one SME coefficient to zero to determine its confidence limit, our method is based on the premise that our null detection does not result from fortuitous cancellations of SME coefficients that hide a signal of oscillation terms in Eq.~(\ref{eq:osc}).  Since the number of SME coefficients is large, this could be an issue.  
This issue was  explicitly addressed in~\cite{paper3} and such a scenario was shown to be quite improbable for the flavor changing coefficients investigated in those previous analyses.  We assume that conclusion holds for this analysis also.

\section{Summary}

We have presented a search for the Lorentz and CPT violating sidereal modulations in the observed neutrino rate predicted by the SME theory for neutrino-antineutrino mixing in the MINOS far detector.   We found no significant evidence for this predicted signal.  When framed in the SME theory~\cite{KM, KM3}, this result leads to the conclusion that we have detected no evidence for Lorentz invariance violation, a result consistent with the analyses in~\cite{ndlv, paper2,paper3}.  We computed upper limits for the 66 SME coefficients appropriate to this analysis.  These limits are the first to be calculated for the SME coefficients governing neutrino-antineutrino mixing. \\

\noindent {\bf Acknowledgements} \\

We gratefully acknowledge our many valuable conversations with Alan Kosteleck\'y and Jorge D\'iaz during the course of this work.  This work was supported in part by the Indiana University Center for Spacetime Symmetries (IUCSS).
We thank the MINOS collaboration for releasing the timestamp information for the neutrino interactions used in this analysis.  We are grateful to the Minnesota Department of Natural Resources, the crew of the Soudan Underground Laboratory, and the staff of Fermilab for their contributions to this effort.

\newpage

\bibliographystyle{elsarticle-num}

\bibliography{lorentz_gH}







\end{document}